\documentclass[letterpaper]{JHEP3}

\usepackage{amsmath}
\usepackage{amssymb}
\usepackage{amsfonts}
\usepackage{feynmf}

\newcommand{\commut}[2]{[\,#1\,,\,#2\,]}

\renewcommand{\bar}{\overline}

\newcommand{\ptm}{\phantom}

\newcommand{\beq}{\begin{eqnarray}}
\newcommand{\eeq}{\end{eqnarray}}

\title{General Massive Spin-2 on de Sitter Background}

\author{Gregory Gabadadze\\Center for Cosmology and Particle Physics, 
Department of Physics\\
New York University, New York, NY 10003 USA
\vspace{0.1cm}\\
Joseph Henry Laboratories, Princeton University\\Princeton, NJ08544, USA\\
E-mail: \email{gg32@nyu.edu}}
\author{Alberto Iglesias\\University of California\\
1 Shields Ave, Davis, CA 95616 USA\\
E-mail: \email{iglesias@physics.ucdavis.edu}}
\author{Yanwen Shang\\University of Toronto\\60 St George Street\\
Toronto, ON M5S 1A7 Canada\\E-mail: \email{ywshang@physics.utoronto.ca}}

\abstract{We study linearised massive gravity on the 
de Sitter background. With a small-enough  graviton mass 
this may have had relevance to inflation, or  
the present-day cosmic acceleration. 
Higuchi has shown that this theory has a ghost 
as long as the dS curvature exceeds the 
graviton mass, $2H^2>m^2$;  this would entail 
rapid instabilities.  In this work we extend the model 
and observe that the helicity-0 mode which is a ghost,
can be turnd into a positive energy state via kinetic  
mixing with the conformal mode. The latter 
gets restricted here by the requirement that the Bianchi 
identities be satisfied.  This eliminates  
the ghost from the linearized theory.  However, the 
spectrum still contains scalar tachyonic instabilities 
with the time scale $\sim 1/H$.   This would have been 
problematic for the early univesre, however, may 
be acceptable  for the present-day accelerated 
expansion as the tachyon instability 
would take the age of the universe to develop.}


\begin{document}

\section{Introduction, discussions and summary}

There are at least two compelling reasons to study theories that modify 
gravity at large distances:
the old cosmological  constant problem \cite{Weinberg} may be solved
in this approach,  and the observed cosmic 
acceleration \cite{Acc} may be explained, see respectively 
\cite {DGP,DGS,ADDG,GS,GGrev} and \cite {Cedric,DDG}. 
Presently,  there is no satisfactory theory of modified gravity 
that could complete the above tasks (see \cite {ADDG,Kaloper,Fbox} and 
\cite {RattazziPorrati,GI,Koyama,Gorbunov,Gregory,
KKTanaka,DGI,Oriol,GGCargese,Myers,Bounce}, for discussions of various 
issues and controversies); the search for a consistent theory continues 
\cite {Toly,GGCargese,Cascade,KoyamaSA,Kakushadze}.

One example of the large distance modified theory  
is four-dimensional massive gravity with cosmologically large graviton 
Compton wavelength. In the linearized approximation around Minkowski 
space this model  is consistent theoretically, but  
fails to  describe the observable world because 
of the vDVZ discontinuity \cite {vDVZ}. Although strong 
non-linear interactions could have cured the discontinuity  
problem \cite {Arkady,DDGV}, the non-linear theory of massive gravitons
by itself is inconsistent \cite  {BD,GGruzinov,CedricRomb,
Nicolis}. 

Certain  Lorentz-violating massive theories 
\cite {Rubakov,Dubovsky,DubovskyIgor} may have better non-linear 
behavior \cite {GGrisa}, as well as interesting phenomenology,  
however, they  exhibit an unusual property of long-range 
instantaneous interactions \cite {GGrisa}.

Furthermore,  the Lorentz-violating ghost condensation 
model that exhibits rich physics \cite {NimaL}, 
as well as  phenomenologically  motivated $f(R)$-type 
models (see, e.g., \cite {FR} and references therein)
have been actively discussed in the literature.

\vspace{0.1in}

The main topic of the present work is massive gravity on the de Sitter 
background \cite {Higuchi}. The interest in this is two-fold:

(A) One could be curious to know what happens to inflation if the graviton 
has a mass $m$ such that $H\gg m$, where $H$ is the Hubble parameter.
Very naively, one may expect that inflation proceeds as  in the 
conventional setup as long as the physical size  of the inflationary 
region is smaller than  the graviton Compton wavelength, while for larger
size the evolution would be altered, perhaps along the lines of 
\cite {ADDG}. 

(B) Some of the known  examples of the 
self-accelerated universe \cite {Cedric,DDG} contain 
in the linearized  approximation  a massive graviton on 
a dS background with  $H\sim m$.  This may well be a  
generic property of a class of self-accelerated solutions, in which case 
understanding of massive gravity on the dS background  may have observationally
interesting consequences for the late time cosmological evolution.

However, before starting to address physically meaningful  
questions on inflation  or late 
time accelerated expansion, one needs to deal with the theoretical 
consistency problems of massive gravity on dS. It has been shown by Higuchi
\cite {Higuchi} that for $2H^2>m^2$ the helicity-0 mode of the 
massive graviton  on the dS background 
becomes a ghost. The ghost would entail very rapid instabilities
of the background.  This eliminates any hope to study 
inflation with massive gravity, as well as  models of 
the late time cosmic acceleration that lie in the parameter range $2H^2>m^2$.

Our aim here is to address precisely these theoretical issues. 
Namely, we will try to modify the linearized 
theory of massive gravity on dS background such that the ghost in  
the case $2H^2>m^2$ is avoided.  We will achieve our goal 
of eliminating the ghost,
however, for   $6H^2> m^2$ we are still left 
with two spin-0 tachyons of negative mass squared $m^2-6H^2$, 
one of which is decoupled from matter sources and the other has 
the coupling strength  measured by the ratio $m^2/(m^2+2H^2)$.  
Is this an improvement over the ghost? 

The answer would depend on a concrete physical 
circumstance at hand. For instance, for the issue (A) above,
if we deal with an inflationary scenario and allow for a small 
graviton mass of the order of the present days Hubble scale 
$m\sim 10^{-33}$eV, then $H\gg m$ and the tachyon instability could 
be very severe; unless further changes are made, it would 
destabilize the inflationary background  in the time scale
$\sim 1/H$, spoiling the inflationary scenario. 
Hence, if graviton is massive, it better acquired 
its mass only during the 
late-time evolution of the universe. 

On the other hand, if we have in mind applications to the 
present-day  accelerated universe as in (B) above, then 
for $2H^2>m^2$ but $H\sim m$ the tachyon instability time is 
of the order of the age of the Universe. In this context, 
replacing the ghost  by  the  tachyon is  a  huge  improvement as 
the ghost would have led to a catastrophic instability. 

It is worth mentioning that in our model we achieve  
a continuous  transition to the massless limit $m\to 0$, as the  
vDVZ  discontinuity will be absent (in a way similar to what happens 
in the AdS case \cite {PorrativDVZ}). Finally, for $m^2>6H^2$ 
our model is also ghost and tachyon free.

\vspace{0.1in}
In the remainder of this section we will try to 
summarize,  in a less cumbersome (but still somewhat technical) 
way, the approach and results of the present work. 

To start with, ghosts present a formidable problem in field theories.  
In the classical limit  they could lead  to unbounded 
negative energy solutions. In the full theory ghosts can 
be quantized  as positive-norm negative energy 
states or, alternatively, as negative-norm positive energy states. 
In the former case they lead to a rapid  vacuum instability via 
a particle-ghost creation process, while in the latter case 
the negative norm states violate unitarity.
Typically, in a theory with a given field content, if ghosts are present,  
there are no tools  to avoid the above problems 
without violating analyticity and causality, or locality 
\cite {Gross}.

Consider for example a scalar ghost. We note that 
a kinetic mixing of two ghosts  may  
eliminate one of them.  To see this we look at 
a Lagrangian which on top of the conventional fields 
contains two additional fields  $\sigma $ and $\tau$,
with kinetic terms
\beq
(\partial_\mu \sigma)^2 +2 \alpha (\partial_\mu \sigma)
(\partial_\mu \tau)+ (\partial_\mu \tau)^2\,, 
\label{sigmatau}
\eeq
where  the parameter $\alpha$ sets the mixing strength.
In the $\alpha \to 0$  limit both $\sigma $ and $\tau$ have 
wrong-sign (ghost-like) kinetic terms (in our conventions of 
the signature $\eta_{\mu\nu}={\rm diag}(-1,1,1,1)$). 
However, due to a large enough mixing, one of the ghosts 
can be turned into a particle by the field redefinition: 
${\sigma} \to   \sigma - \alpha \tau$. The resulting Lagrangian 
takes the form:
\beq
(\partial_\mu {\sigma})^2   - [\alpha^2 -1] (\partial_\mu \tau)^2\,.
\label{gh}
\eeq
For  $|\alpha|>1$ the kinetic term of $\tau$
flips the sign to the ``right one''\footnote{In 
the space of the mixing  parameter,
$\alpha=1$ is a singular point where interactions 
with other fields would in general become infinitely strong.
The two theories with different signs of the kinetic term of $\tau$
are separated by this singular point.}. 

Note that,  if the kinetic term of $\sigma$ had an 
opposite sign in (\ref {sigmatau}) (i.e., if it were  
not ghost-like)  one would not be able  to flip the ghost-like 
sign of the  $\tau$ kinetic term via the diagonalization: 
It takes a ghost to kill a ghost!
  
Is the above exercise meaningful? After all we are still  
left with one ghost ${\sigma}$ that is bad-enough to give 
rise to all the known ghost-related  problems.

The answer to the above question would be positive if the 
${\sigma}$  field  is constrained further,  so that 
in the end,  this field is left  non-dynamical.  
We will argue below,  and show in the text, 
that such a  mechanism can be at work in  
a theory of massive gravity on the dS background with an 
additional scalar (the latter can be set to decouple from the matter 
stress-tensor). 

In this work we will be discussing  the Pauli-Fierz (PF) 
mass term,  which has a virtue of being 
ghost-free  in the flat space limit:
\beq
\mathcal L_\textrm{PF}=-{m^2\over 4} (h_{\mu\nu}^2 - h^2)\,.
\label{PF}
\eeq
Here the indices are contracted via the background dS metric 
$\gamma_{\mu\nu}$ and its inverse.    Let us
look at the decomposition of the metric perturbation on dS space,  
in terms of  the transverse-traceless tensor $h^{TT}_{\mu\nu}$, 
transverse  vector  $ V_\mu^T$, conformal scalar $\sigma$ and 
longitudinal scalar $\tau$:
\beq
h_{\mu\nu} = h^{TT}_{\mu\nu}+ \nabla_\mu V_\nu^T+ 
\nabla_\nu V_\mu^T +\gamma_{\mu\nu} \sigma + \nabla_\mu  
\nabla_\nu \tau\,.
\label{hTTIntro}
\eeq
Unlike in Einstein's gravity where the  $\tau$ field is gauge 
removable, in the massive theory it acquires physical meaning
of the helicity-0 state of the massive spin-2.

The Bianchi identities combined together with 
the equations of motion following from the massive theory with 
the  term (\ref{PF}) necessarily imply that 
\beq
\nabla^\mu \nabla^\nu h_{\mu\nu} = \square h\,.
\label{Bianchi}
\eeq
Using (\ref {hTTIntro}) and (\ref {Bianchi}) we will get 
\beq
\square \sigma= H^2 \square \tau\,.
\label{constr}
\eeq
Although there will be kinetic and mixing terms for 
$\sigma $ in the Lagrangian, the $\sigma $ field  
in the end should be supplemented by the above constraint 
which expresses it via $\tau$ and excludes it 
from the counting of the physical degrees of freedom.

Let us step back for a moment and return to the Lagrangian in which  
the constraint (\ref {constr}) has not been enforced yet.
The relevant properties for us are encoded in the 
conformal mode $\sigma$  and helicity-0 state described by 
$\tau$, so we focus on these two fields. The $\tau$
field does not enter the EH Lagrangian, but enters 
the PF terms (\ref {PF}) in two different ways. 
First it acquires a kinetic mixing term with  $\sigma$, and second
it gets its own kinetic term due to the fact that the background is 
non-trivial  (the covariant derivative does not commute with  
the dS space d'Alambertian, while it does  so in Minkowski space).  
The kinetic term for  $\tau$ arising  
from  the  PF terms (\ref {PF}) is (dropping the overall factor of 
$3/4$ here and below in this section):
\beq
 m^2 H^2 (\nabla_\mu \tau)^2\,.
\label{kintau}
\eeq
This is a wrong-sign (ghost-like) kinetic term. Hence, if $\tau$ 
had no mixing with other fields, it would be a ghost. In the present  
case, $\tau$ does mix with $\sigma$, and the latter by itself has a 
ghost-like kinetic term  that arises from the EH Lagrangian.  
We can diagonalize the  $\sigma$ - $\tau$ kinetic terms 
by the shift  $\sigma = {\bar \sigma} 
+ (m^2/2) \tau$ (leaving the mass terms mixed since 
this is not important here).  The result reads:
\beq
2 (\nabla_\mu {\bar \sigma})^2 - {m^2\over 2} (m^2 -2 H^2) 
(\nabla_\mu  \tau)^2 \,.
\label{kintaufin}
\eeq 
As we can see, if $m^2>2H^2$, the helicity-0 mode $\tau$ acquires
a positive-sign kinetic term.  
When $m^2 =2H^2$  its kinetic term disappears (see discussions and 
references in \cite {Deser,Gabadadze:2008uc}). But when  $2H^2>m^2 $,  
the helicity-0 mode becomes  
a ghost\footnote{Note that the same conclusions will be reached if 
${\bar \sigma}$ is expressed from (\ref {constr}) in terms of 
$\tau$ and substituted into (\ref {kintaufin}).}.
The kinetic term of ${\bar \sigma}$ remains ghost-like all the time.
However, as before ${\bar \sigma}$ is not an independent 
dynamical field as it  gets related to $\tau$ via 
the constraint (\ref {constr}). 
Hence, for $2H^2>m^2$, we are left with two fields 
with the ghost-like kinetic terms in the Lagrangian, and
a constraint that relates them, which cannot help to circumvent any of
the problems caused by the ghost.  This is a
convenient way of capturing some of 
the key results of Higuchi \cite {Higuchi}.

The main  idea of this paper is to turn the $\tau$ field into a 
field with a right-sign kinetic term via additional mixing between 
${\bar \sigma}$ and $\tau$, in analogy with the scalar model 
(\ref {sigmatau}). For this we introduce, as in \cite{Gabadadze:2008uc},
a scalar field $\phi$ which facilitates a new kinetic mixing between 
${\bar \sigma}$ and $\tau$ in (\ref {kintaufin}). 
This mixing is designed such that 
after the diagonalization  of the kinetic terms 
the $\tau$  field acquires an additional 
right-sign contribution and turns into a field with a right-sign kinetic term 
even for $2H^2>m^2$. The ${\bar \sigma}$ retains its ghost-like 
nature, which again is harmless, since this field is constrained.

One could integrate out the additional scalar $\phi$, in which case, 
one would be left with a Lagrangian in which both the 
EH and PF mass terms are modified.  For convenience we will retain  
the  scalar  $\phi$, since the Lagrangian is manifestly local in this case.

In  the rest of the paper  we will work with 
general expressions without separating conformal and 
helicity-0 modes, although our  results may be more conveniently
understood as described above.

What is left out of the present work is the discussion of a nonlinear 
completion of the models that we are discussing here. One such possibility
could be the DGP model \cite {DGP} endowed with an 
additional scalar dynamics, that at the linearized level would reduce to 
the theory presented in this article.  Related to the previous comment, we 
will also not discuss any  implications of potential  quantum-loop 
corrections -- there are challenging issues to be understood 
already at the classical level in the linearized theory.

\section{Action and Equations of Motion}

We start by considering a quadratic action for a graviton of mass 
$m$ and a scalar field with kinetic mixing on de Sitter space, 
coupled to a conserved matter stress-tensor:
\begin{equation}
\label{eq:L0}
\mathcal{L}_{\textrm{eff}}=\mathcal{L}^{(2)}_{\textrm{EH}} (h_{\mu\nu})
-\frac{1}{4}m^2(h_{\mu\nu}^2-h^2)
-\phi \mathcal{O}^{\mu\nu} h_{\mu\nu}
+\phi \mathcal{K}\phi+ h_{\mu\nu}T^{\mu\nu}+q\phi T~,
\end{equation}
where $\mathcal{L}_\textrm{EH}^{(2)}$ is the second order 
expansion of the Einstein-Hilbert
action around de Sitter space with the 
cosmological constant $\Lambda=3H^2$, and the operator 
\begin{equation}
\label{eq:def_O}
\mathcal O_{\mu\nu}=\nabla_\mu\nabla_\nu -\gamma_{\mu\nu}\square
-3 H^2\gamma_{\mu\nu}~.
\end{equation}
We have introduced a coefficient $q$ which is assumed to be a constant 
whose value is chosen later. 
This form of the operator $\mathcal O_{\mu\nu}$ is motivated by its
transversality on the de Sitter background. For future
convenience, we define 
\begin{equation}
Q\equiv -\square-4H^2~.
\end{equation}
The operator $\mathcal K$ that appears in \eqref{eq:L0} also remains 
undetermined at this stage. We shall 
find that a particular choice of this operator gives rise to special
simplifications of the theory.  At this point we just assume that $\mathcal K$ 
is a scalar operator and contains at most second derivatives. 
Hence, it takes the following form
\begin{equation}
\mathcal K= A\,\square +B~.
\end{equation}
Since  $A$ and $B$ are assumed to be constants, then 
$\mathcal K$ commutes with $Q$.

There are both kinetic and mass mixings between graviton and scalar in 
(\ref{eq:L0}), and for non-vanishing value of the parameter $q$ both fields 
are sourced by matter.  The equations of motion resulting from this action are 
\beq
\label{eq:eom}
G^{\textrm{ds}}_{\mu\nu}-\frac{m^2}{2}(h_{\mu\nu}-\gamma_{\mu\nu} h) 
-\mathcal O_{\mu\nu}\phi
&=&-T_{\mu\nu}~,\\
\label{eq:eom_phi}
 \mathcal O^{\mu\nu} h_{\mu\nu}-2\mathcal K\phi&=&q T~.
\eeq
As in the case of the pure PF gravity, the Bianchi identities give rise to 
the following relations
\begin{equation}
\nabla^\mu h_{\mu\nu}=\nabla_\nu h~,
\end{equation}
which can be used to reduce the equations of motion to the system:
\begin{equation}
\frac{1}{2}\left[\square h_{\mu\nu}-(2H^2+m^2) h_{\mu\nu}
-\gamma_{\mu\nu}(H^2-m^2)h-\nabla_\mu\nabla_\nu h\right]=
-T_{\mu\nu}+\mathcal O_{\mu\nu} \phi~,
\end{equation}
with the trace equation being
\begin{equation}
\label{eq:trace_einstein}
(3H^2-\frac{3}{2} m^2)h+3 Q \phi=T~.
\end{equation}
The field $h_{\mu\nu}$ is not  traceless, therefore, to derive its
propagator, one must use the Lichnerowicz operator $\Delta_L$ defined in 
the Appendix by equation 
\eqref{eq:lich}, with which one can write the equations of motion 
(\ref{eq:eom}) as
\begin{equation}
\frac{1}{2}(\Delta_L-6H^2+ m^2) h_{\mu\nu}
=T_{\mu\nu}-
\mathcal O_{\mu\nu}\phi-\frac{1}{2}\left[(3H^2- m^2)\gamma_{\mu\nu}
+\nabla_\mu\nabla_\nu\right]h~,
\end{equation}
or, by defining 
\begin{equation}
\mathcal M_{\mu\nu}\equiv (3H^2- m^2)\gamma_{\mu\nu}+\nabla_\mu\nabla_\nu,
\end{equation}
as
\begin{equation}
\frac{1}{2}(\Delta_L-6H^2+ m^2) h_{\mu\nu}
=T_{\mu\nu}-\mathcal O_{\mu\nu}\phi-\frac{1}{2} \mathcal M_{\mu\nu}h.
\end{equation}

Finally, the equation of motion for $\phi$, \eqref{eq:eom_phi}, implies that
\begin{equation}
-3H^2 h-2\mathcal K\phi=qT~,
\end{equation}
and that, together with equation \eqref{eq:trace_einstein}, yields 
the following solutions for $\phi$ and $h$:
\beq
\label{eq:solution_phi_h}
\phi&=&\frac{(1+q)H^2-\frac{1}{2} q m^2}{3 H^2 Q+ (m^2-2H^2) \mathcal K} T, \\
h&=&\frac{ q Q+\frac{2}{3}\mathcal K}{\frac{1}{2} q m^2-(1+q)H^2}\phi.
\eeq
It can be easily checked that if we set $ m^2=2H^2$, we recover 
the results for both $\phi$ and $h$ studied in \cite{Gabadadze:2008uc}.

\section{Solutions}

The goal here is to obtain a theory free of ghosts.  Our approach will 
be similar to the one adopted in  \cite{Gabadadze:2008uc}, in which the 
parameters of the original action are chosen so that 
there are no ghost-like poles in the propagators of the physical fields.

We define the physical metric perturbation 
\beq\label{shift}
h_{\mu\nu}^\textrm{phy}=h_{\mu\nu}-q\gamma_{\mu\nu}\phi~, 
\eeq
that captures  the entire response of the system to 
$T_{\mu\nu}$ (after performing this shift, 
the scalar $\phi$ is no longer sourced by $T$).
 The value of the field is given by:
\beq
\label{eq:master}
\frac{1}{2}h^{\textrm{phy}}_{\mu\nu}
&=&\frac{1}{\Delta_L-6H^2+ m^2}\left\{T_{\mu\nu}-\mathcal O_{\mu\nu}\phi
-\frac{1}{2}\mathcal M_{\mu\nu} h\right\}+\frac{1}{2}\gamma_{\mu\nu}q\phi\\
&=&\frac{1}{\Delta_L-6H^2+ m^2}\left\{T_{\mu\nu}-\mathcal O_{\mu\nu}\phi
-\frac{1}{2}\mathcal M_{\mu\nu} h
+\frac{q(Q-2H^2+m^2)}{2}\gamma_{\mu\nu}\phi\right\}~,
\eeq
where $q$ and $\mathcal K$ are still to be determined. 
Notice that generically $\phi$ and $h$ contain single poles.

Following the strategy of \cite{Gabadadze:2008uc} to 
cancel all the unwanted poles, we first  
focus on the terms inside the curly brackets on the r.h.s. 
of equation \eqref{eq:master}. There, the single poles are 
carried by the two terms including $\phi$ and $h$ respectively.
Their pole parts  must cancel each other out.  The finite 
remnant of this cancellation, once taken out of the curly brackets,  
turns into a single pole term that eventually can be canceled 
by $\gamma_{\mu\nu}q\phi$.  The terms proportional to 
$\nabla_\mu\nabla_\nu T$ are harmless, at least 
at the tree-level, since they lead to vanishing contributions 
when contracted with a conserved $T_{\mu\nu}$.

Notice that there is nothing we can do about the operator 
$\mathcal M_{\mu\nu}$ 
in front of $h$. Therefore for any cancellation to become possible, the term
$\mathcal O_{\mu\nu}\phi$ must somehow contain a term of the form of 
$\mathcal M_{\mu\nu}h$
(or $\mathcal M_{\mu\nu}\phi$ since $\phi$ and $h$ are related by equation 
\eqref{eq:solution_phi_h}).  The solution is then to demand that
\begin{equation}
\mathcal O_{\mu\nu}\phi=\frac{1}{3}
\gamma_{\mu\nu}T+\mathcal M_{\mu\nu}\phi~.
\end{equation}
The coefficient in front of $\mathcal M_{\mu\nu}$ is fixed to be 
one\footnote{One may attempt to
change the coefficient of the $\nabla_\mu\nabla_\nu$ term in the definition of 
$\mathcal O$, but this only amounts to a rescaling of $\phi$, $q$ and 
$\mathcal K$, which are at this stage not fixed.}.
This leads to
\begin{equation}
(Q-2H^2+ m^2)\frac{(1+q)H^2-\frac{1}{2}q m^2}{3H^2 Q-(2H^2- m^2)\mathcal K}
=\frac{1}{3}~,
\end{equation}
and we find
\begin{equation}
\label{eq:def_tilde_m}
(2H^2- m^2)\mathcal K=3q(\frac{1}{2} m^2-H^2)Q+6(H^2-\frac{1}{2} m^2)
\left((1+q)H^2-\frac{1}{2} q m^2\right)~.
\end{equation}
Notice that for  $ m^2=2H^2$, the above equation is automatically satisfied 
and $\mathcal K$ remains completely arbitrary. That is why for this
special case one finds additional freedom 
as discussed in \cite{Gabadadze:2008uc}.

Things are very different when $2H^2>m^2$. In this case, the 
form of $\mathcal K$ is fixed to be
\begin{equation}
\mathcal K=-\frac{3q}{2} Q+ 3\left((1+q)H^2-\frac{1}{2}q m^2\right)~.
\end{equation}
Such a choice of $\mathcal K$ leads to an immediate 
consequence that $h$ is directly proportional
to $\phi$. Indeed, we have
\begin{equation}
h=-2\phi~.
\end{equation}
Therefore the single poles inside the curly brackets, 
besides the $q(Q-2H^2+ m^2)\phi$
term, cancel exactly. Furthermore, we find
\beq
\phi=-\frac{1}{3(\square+6H^2-m^2)} T~,
\eeq
which in its turn implies that
\begin{equation}\label{hphy}
\frac{1}{2}h^{\textrm{phy}}_{\mu\nu}=\frac{1}{\Delta_L-6H^2+m^2}
\left( T^{(1/2)}_{\mu\nu}+\frac{1+q}{6} \gamma_{\mu\nu} T\right)~.
\end{equation}

Recall now that $h_{\mu\nu}=h_{\mu\nu}^\textrm{phy}-q\gamma_{\mu\nu}\phi$, 
and if we rewrite the original action in terms of the ``physical''
metric perturbation, we get
\beq\label{leff}
{\cal L}_{eff}&=&{\cal L}^{(2)}_{EH+PF}(h_{\mu\nu}^{\textrm {phy}})
-(1+q)\phi {\cal O}^{\mu\nu}h_{\mu\nu}^{\textrm {phy}}
-\frac{3}{2}q(q+1)\phi\Box\phi-\frac{3}{2}
q m^2h^{\textrm {phy}}\phi+\nonumber\\
&&+\frac{3}{2}(2q-1)(q m^2-2(q+1)H^2)\phi^2+
h_{\mu\nu}^{\textrm {phy}}T^{\mu\nu}~.
\eeq
The theory described by (\ref{leff}) 
contains six propagating degrees of freedom: 
five polarizations of a massive spin two,  plus the extra 
scalar that we introduced in (\ref{eq:L0}). 
After the shift that defines $h_{\mu\nu}^\textrm{phy}$,
$T_{\mu\nu}$ only sources five degrees of freedom --  
the polarizations of $h_{\mu\nu}^\textrm{phy}$.
Therefore, if we neglect for a moment the mixing 
between $h_{\mu\nu}^\textrm{phy}$ and $\phi$ we get 
that the former propagates  five degrees of freedom, 
two of which do not couple to a conserved 
source, giving rise to the usual propagator of massive gravity on dS.
A single graviton exchange between two sources leads 
in this case to a diagram like the one of figure 1a.

\setlength{\unitlength}{1mm}
\begin{figure}\label{fig1}
\begin{fmffile}{figure1}
$a$.
	\begin{fmfgraph*}(60,20)
	\fmfleft{i1,i2}
	\fmfright{o1,o2}
	\fmf{plain}{i1,v1,i2}
	\fmf{plain}{o1,v2,o2}
    	\fmf{wiggly, label=$h^\textrm{phy}$}{v1,v2}
	\fmfdotn{v}{2}
	\end{fmfgraph*}
$b$.
	\begin{fmfgraph*}(80,20)
	\fmfleft{i1,i2}
	\fmfright{o1,o2}
	\fmf{plain}{i1,v1,i2}
	\fmf{plain}{o1,v4,o2}
    	\fmf{wiggly, label=$h^\textrm{phy}$}{v1,v2}
	\fmf{dashes,label=$\phi$}{v2,v3}
	\fmf{wiggly, label=$h^\textrm{phy}$}{v3,v4}
	\fmfdotn{v}{4}
	\end{fmfgraph*}
\end{fmffile}
\caption{$a$. single graviton (wiggly line) exchange with no scalar mixing.
$b$. Case with scalar (dashed line) mixing. 
The general case would contain a chain 
alternating scalar and graviton.  }
\end{figure}
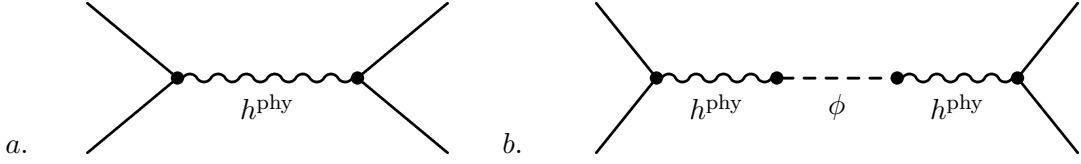

Including the mixing terms perturbatively, would give diagrams of 
the type depicted in figure 
1b,  in which a combination of the helicity-$0$ component of 
$h_{\mu\nu}^\textrm{phy}$  
oscillates into the scalar $\phi$ and back again. But at every 
instant it is only one scalar degree
of freedom that is propagating between the sources,  
on top of the ordinary helicity-2 components. 

The resummation of all such tree-level  
diagrams gives the propagator (\ref{hphy}).
In the latter the contributions of the two helicity-2 states 
are put  together in  the term proportional 
to $T_{\mu\nu}^{1/2}$,  while the contribution of 
the single scalar mode (which is a superposition  
of the helicity-0 mode of $h_{\mu\nu}^\textrm{phy}$ and $\phi$)  
is in the term proportional to $T$. Notice that the 
latter has positive residue  as long as $q>-1$. 

In order to determine whether the remaining sixth degree 
of freedom -- the other combination of the  helicity-0 mode and 
$\phi$,   which is not  sourced by $T_{\mu\nu}$ --  
is a ghost or not  we use the following trick: 
we temporarily  set to zero $T_{\mu\nu}$ and include a 
putative source $J$ via the term  $+J\phi$ in the action.  
$J$ excites different combinations of the helicity-0 mode 
and $\phi$; then,  if there is a  ghost-like  excitation in the 
sixth mode that's not sourced by $T$, it would be sourced by $J$. 
Performing this  analysis, we find that for
\beq\label{qscal}
q>\frac{2H^2}{m^2-2H^2}~,
\eeq
no ghosts are excited by $J$ either. Hence, this linearized  theory is 
free of ghosts. Note also that for $2H^2>m^2$ the condition 
(\ref{qscal}) is less restrictive than the one for the other scalar 
combination ($q>-1$).

Another point to be mentioned is that both the field $\phi$ (for $q$ 
satisfying (\ref{qscal})) and the scalar component of 
$h^\textrm{phy}_{\mu\nu}$ have positive mass squared for $m^2>6H^2$, while 
they become tachyonic when $m^2<6H^2$ (see, also eq. 
(\ref {pro}) below. One of these tachyons is 
decoupled from the matter stress-tensor 
and another one couples to it with the strength
that is suppressed as $m^2/(m^2+2H^2)$.  The cases where this may
not lead to a problem were discussed in Section 1. 
In particular, such a  tachyon is  
a big improvement over the ghost, as for $2H^2>m^2$ but  
$H\sim m$ the tachyon instability time is of the order of the age 
of the Universe, while in the case of a ghost the instability would 
have been catastrophically fast.

In our model there is 
a continuous transition to the massless limit $m\to 0$ and the
vDVZ  discontinuity is absent for some choices of $q$, one of which
we discuss in the next section.

\section{An example with no vDVZ discontinuity}

At this point we make a choice for the value of the parameter $q$. We take
\beq\label{q}
q= \frac{m^2 -2H^2}{m^2+2H^2}~,
\eeq 
and that renders the theory 
ghost free for all positive values of $m^2$ and $H^2$.

Indeed, with the choice (\ref{q}) we find that the physical graviton 
field has  the following structure:
\beq\label{pro}
\frac{1}{2}h^{\textrm{phy}}_{\mu\nu}=\frac{1}
{\Delta_L-6H^2+m^2}T^{(1/2)}_{\mu\nu}
+\gamma_{\mu\nu}\frac{1}{3}\frac{m^2}{m^2+2H^2}\frac{1}{-\Box-6H^2+m^2}T~,
\eeq
the first term describes the propagation of two tensor polarizations 
of squared mass $m^2$,  while the second one shows the propagation of a 
scalar component with a positive residue $m^2/3(m^2+2H^2)$.

A scan of the possible values of $H^2$ in  (\ref{q}) is also interesting:
For $2H^2=m^2$ we have $q=0$ which coincides with the special massive spin-2 
theory described in \cite{Gabadadze:2008uc}.
 
For $H^2\gg m^2$, making an expansion in powers 
of $\epsilon=\sqrt{m^2/2H^2}$ in 
(\ref{leff}), we get in the  first order:
\beq
{\cal L}_{eff}&=&{\cal L}^{(2)}_{EH+PF}(h_{\mu\nu}^{\textrm {phy}})
+\frac{1}{2}\varphi(\Box+9H^2)\varphi
+h_{\mu\nu}^{\textrm {phy}}T^{\mu\nu}
-\sqrt{\frac{2}{3}}\epsilon
\varphi({\cal O}^{\mu\nu}-\frac{3}{2}H^2)h_{\mu\nu}^{\textrm {phy}}~,
\eeq
where we introduced the canonically normalized field 
$\varphi=\sqrt{3m^2/4H^2}\phi$.
If we keep $H^2$ finite in the $\epsilon\to 0$ limit (i.e., a massless limit)
the mixing disappears between $\varphi$ 
and $h_{\mu\nu}^{\textrm {phy}}$, so does the PF term 
(proportional to $H^2\epsilon^2$). 
Thus, the theory becomes GR plus a decoupled scalar $\varphi$.
Furthermore, the limit is smooth, since from (\ref{pro}) we can see the 
second term becoming negligible leaving only the two polarizations of 
a massless graviton.

\section*{Acknowledgments}

We would like to thank  Stefan Hofmann, Slava Mukhanov and Massimo Porrati 
for useful discussions.  GG thanks Igor Klebanov and Physics Department 
of Princeton University for hospitality where a part of this work was done.
AI and YS would like to thank the members of the Center for Cosmology and
Particle Physics of New York University for their hospitality while this
work was being completed.  The work of GG was supported by the 
NSF (PHY-0758032), and NASA ( NNGG05GH34G), AI was supported by DOE Grant 
DE-FG03-91ER40674.

\appendix

\section{Conventions and definitions}
\label{convention}


On the de Sitter background the Riemann tensor is given by
\begin{equation}
\label{eq:desitter_R}
R_{\mu\nu\rho\sigma}=H^2(\gamma_{\mu\rho} \gamma_{\nu\sigma}
		-\gamma_{\mu\sigma} \gamma_{\nu\rho}),
\end{equation}
therefore the Ricci tensor takes the form
\begin{equation}
R_{\mu\nu}=3 H^2 \gamma_{\mu\nu}\,, 
\end{equation}
and the Ricci  curvature scalar equals to $R=12 H^2$.
Furthermore,
\begin{equation}
\begin{split}
\square\, \nabla_\mu\nabla_\nu\varphi
=&(\nabla^\rho\nabla_\mu\nabla_\rho\nabla_\nu
+\nabla^\rho\commut{\nabla_\rho}{\nabla_\mu}\nabla_\nu)\varphi\\
=&(\nabla_\mu\square\,\nabla_\nu
+\nabla^\rho\commut{\nabla_\rho}{\nabla_\mu}\nabla_\nu
+\commut{\nabla^\rho}{\nabla_\mu}\nabla_\rho\nabla_\nu)\varphi\\
=&(\nabla_\mu\nabla_\nu\,\square
+\nabla^\rho\commut{\nabla_\rho}{\nabla_\mu}\nabla_\nu
+\commut{\nabla^\rho}{\nabla_\mu}\nabla_\rho\nabla_\nu
+\nabla_\mu\commut{\nabla_\rho}{\nabla_\nu}\nabla^\rho)\varphi\\
=&(\nabla_\mu\nabla_\nu\,\square
-\nabla^\rho R^\lambda_{\phantom{\lambda}\nu\rho\mu}\nabla_\lambda
-R^{\lambda\phantom{\rho}\rho}_{\phantom{\lambda}\rho\phantom{\rho}\mu}
	\nabla_\lambda\nabla_\nu
-R^{\lambda\phantom{\nu}\rho}_{\phantom{\lambda}\nu\phantom{\rho}\mu}
	\nabla_\rho\nabla_\lambda
-\nabla_\mu R^{\lambda\rho}_{\phantom{\lambda\rho}\rho\nu}
	\nabla_\lambda)\varphi\,.
\end{split}
\end{equation}
Using \eqref{eq:desitter_R} we get
\begin{equation}
(\square\,\nabla_\mu\nabla_\nu-
\nabla_\mu\nabla_\nu\,\square)\varphi=
8H^2\left(\nabla_\mu\nabla_\nu-\frac{1}{4}
\gamma_{\mu\nu}\square\right)\varphi\,.
\end{equation}
The Lichnerowicz operator acting on a general rank-2 tension is given by
\begin{equation}
\Delta_L h_{\mu\nu}=-\square h_{\mu\nu}+
2 R^{\rho}_{\phantom{\rho}\mu\nu\sigma} h^{\sigma}_{\phantom{\sigma}\rho}
+R_{\rho\mu} h^{\rho}_{\ptm{\rho}\nu}+R_{\rho\nu} h^{\rho}_{\ptm{\rho}\mu}\,.
\end{equation}
Using the Riemann tensor given in (\ref {eq:desitter_R}), 
this becomes
\begin{equation}
\label{eq:lich}
\begin{split}
\Delta_L h_{\mu\nu}=&-\square h_{\mu\nu} +2 H^2(h_{\mu\nu}-\gamma_{\mu\nu} h)
	+6 H^2 h_{\mu\nu}\\
=&-\square h_{\mu\nu}+8 H^2 h_{\mu\nu}-2 H^2 \gamma_{\mu\nu} h~.
\end{split}
\end{equation}
Define
\begin{equation}
P_{\mu\nu}=\nabla_\mu\nabla_\nu-\frac{1}{4}\gamma_{\mu\nu}\square\,,\;
Q=-\square-4 H^2\,,\;
S=-\square+4 H^2\,.
\end{equation}
It is easy to verify that 
\begin{equation}
\Delta_L P_{\mu\nu}\varphi=(-\square+8H^2) 
P_{\mu\nu}\varphi=-P_{\mu\nu}\,\square\varphi~,
\end{equation}
and therefore
\begin{equation}
(\Delta_L-4H^2) P_{\mu\nu}\varphi=P_{\mu\nu} Q\varphi.
\end{equation}
Similarly
\begin{equation}
\Delta_L \gamma_{\mu\nu}\varphi=-\square \gamma_{\mu\nu}\phi.
\end{equation}

Some other useful identities are:
\begin{gather}
\commut{\nabla_\mu}{\nabla_\nu}\varphi=0~,\\
\commut{\square}{\nabla_\mu}\varphi
=(\nabla^\nu\nabla_\nu\nabla_\mu-\nabla^\mu\nabla^\nu\nabla_\nu)\varphi
=\commut{\nabla^\nu}{\nabla_\mu}\nabla_\nu\varphi=3H^2\nabla_\mu\varphi~,
\end{gather}
and following the second identity given above, we also find
\begin{equation}
\nabla^\mu\square\nabla_\mu\varphi=\square^2\varphi+3H^2\square \varphi.
\end{equation}
As a consequence of this, the operator $\mathcal O^{\mu\nu}$ annihilates
$\nabla_\mu\nabla_\nu\varphi$ for any scalar $\varphi$, since
\begin{equation}
\begin{split}
\mathcal O^{\mu\nu}\nabla_\mu\nabla_\nu\varphi
=&\nabla^\mu\nabla^\nu\nabla_\mu\nabla_\nu\varphi-
\square^2\varphi-3H^2\square\varphi\\
=&\nabla^\mu\square\nabla_\mu\varphi-\square^2\varphi-3H^2\square \varphi\\
=&\square^2\varphi+3H^2\square\varphi-\square^2\varphi-3H^2\square\varphi=0.
\end{split}
\end{equation}
This is consistent with the linearized perturbations of the Ricci scalar
on the dS background being proportional to $\mathcal O^{\mu\nu}h_{\mu\nu}$.

\section{Some more discussions}

Let us perform the following transformation 
\begin{equation}
\label{eq:more_transformation}
h^\textrm{phy}_{\mu\nu}=\tilde h_{\mu\nu}+c\nabla_\mu\nabla_\nu\phi,
\end{equation}
where $c$ is a constant to be determined.

In what follows we discuss how the effective Lagrangian 
\eqref{leff} varies
under such a transformation. Clearly 
$\mathcal L^{(2)}_\textrm{EH}(\tilde h_{\mu\nu})
=\mathcal L^{(2)}_\textrm{EH}(h^\textrm{phy}_{\mu\nu})$, 
because what
we have done is just a gauge transformation 
for $h^\textrm{phy}_{\mu\nu}$ in the
case of pure gravity.

As mentioned above, $\mathcal O^{\mu\nu}$ 
annihilates $\nabla_\mu\nabla_\nu\phi$,
therefore, $\phi\mathcal O^{\mu\nu} 
h^\text{phy}_{\mu\nu}=\phi\mathcal O^{\mu\nu}\tilde h_{\mu\nu}$
is also invariant. The term $\mathcal L_\textrm{PF}$ of course breaks 
the gauge symmetry and it varies:
\begin{equation}
\begin{split}
\mathcal L_\textrm{PF}(h^\textrm{phy}_{\mu\nu})
=&-\frac{1}{4}m^2(h^{\textrm{phy}\,2}_{\mu\nu}-h^{\textrm{phy}\, 2})\\
=&\mathcal L_\textrm{PF}(\tilde h_{\mu\nu})\\
&-\frac{1}{2}m^2c\tilde h_{\mu\nu}\nabla^\mu\nabla^\nu\phi
-\frac{1}{4}m^2c^2\phi\nabla^\mu\square\nabla_\mu\phi
+\frac{1}{2}m^2c\tilde h\square \phi+\frac{1}{4}m^2 c^2\phi\square^2\phi\\
=&\mathcal L_\textrm{PF}(\tilde h_{\mu\nu})
-\frac{1}{2}m^2 c \tilde h_{\mu\nu}\nabla^\mu\nabla^\nu\phi
-\frac{3}{4}m^2 H^2 c^2\phi\square\phi
+\frac{1}{2}m^2 c \tilde h\square\phi,
\end{split}
\end{equation}
where we have used some of the identities for swapping operators given 
in  Appendix \ref{convention}.
The only other term in \eqref{leff} that is not invariant is
\begin{equation}
-\frac{3}{2}qm^2h^\textrm{phy}\phi=-\frac{3}{2}qm^2 \tilde h\phi
-\frac{3}{2}qcm^2\phi\square\phi.
\end{equation}
Therefore we find
\begin{equation}
\begin{split}
\mathcal L^{(2)}_\textrm{eff}=
&\mathcal L^{(2)}_\textrm{EH+PF}(\tilde h_{\mu\nu})
-\left[\frac{1}{2}m^2 c+(1+q)\right](\phi\nabla_\mu\nabla_\nu\tilde h_{\mu\nu}
-\phi\square\tilde h)\\
&+\left[3(1+q)H^2-\frac{3}{2}qm^2\right]\tilde h\phi
-\left[\frac{3}{2}q(q+1+m^2 c)+\frac{3}{4}m^2H^2 c^2\right]\phi\square\phi\\
&+\frac{3}{2}(2q-1)\left[qm^2-2(q+1)H^2\right]
\phi^2+\tilde h_{\mu\nu}T^{\mu\nu}.
\end{split}
\end{equation}
It is not completely trivial that one can 
now set the value of a single constant $c$
to remove all the derivative mixings between $\tilde h$ and $\phi$.
To do so we must choose
\begin{equation}
c=-\frac{2(1+q)}{m^2},
\end{equation}
in which case the theory reduces to
\begin{equation}
\label{eq:new_Leff}
\begin{split}
\mathcal L^{(2)}_\textrm{eff}=&\mathcal 
L^{(2)}_\textrm{EH+PF}(\tilde h_{\mu\nu})
-\frac{3}{2}\big[qm^2-2(1+q)H^2\big]\tilde h\phi
+\frac{3(q+1)}{2m^2}\big[qm^2-2(1+q)H^2\big]\phi\square\phi\\
&+\frac{3}{2}(2q-1)
\left[qm^2-2(q+1)H^2\right]\phi^2+\tilde h_{\mu\nu}T^{\mu\nu}.
\end{split}
\end{equation}
Here we find a curious result. All the terms that involve the scalar
$\phi$ contain a common factor. To remove the mixing between $\tilde h$ and
$\phi$, one must set this factor to zero by choosing 
\begin{equation}
\label{eq:qH}
q=\frac{2H^2}{m^2-2H^2}.
\end{equation}
In this case, all the terms that involve $\phi$ disappear simultaneously,
and the theory becomes just a conventional massive spin-2 on
de Sitter background without any additional degrees of freedom! 
It means that with this choice of $q$ and the special form of operator 
$\mathcal O^{\mu\nu}$
and $\mathcal K$, the initial Lagrangian \eqref{eq:L0} is nothing but 
a pure linearized gravity on de 
Sitter background, but expressed after a certain 
conformal and gauge transformations have been performed.  This was shown
by Higuchi \cite{Higuchi} to have a ghost for $m^2<2H^2$. Notice that our
consistency region for $q$ \eqref{qscal} is just right 
above the value \eqref{eq:qH}.

\end{document}